\documentclass{article}
\usepackage{spconf,amsmath,graphicx,hyperref}

\usepackage{amsmath,amssymb}
\usepackage{booktabs}
\usepackage{cite}
\usepackage{xcolor}
\usepackage{tikz}
\usepackage{microtype}
\usetikzlibrary{arrows.meta,positioning}

\title{SEMAC: Spherical Energy Matching Ambisonics Codec\\for Efficient Higher-Order Ambisonics Compression}

\name{Christoph Hold$^{1,2}$, Jan Buethe$^{1}$, Jean-Marc Valin$^{3}$, Jan Skoglund$^{3}$, Toni Hirvonen$^{4}$, Nils Peters$^{5}$}
\address{$^{1}$Meta \enspace
    $^{2}$Reality Labs Research \enspace
    $^{3}$Google \enspace
    $^{4}$Samsung Research America \enspace
    $^{5}$Trinity College Dublin}

\begin{document}

\maketitle

\begin{abstract}
Higher-order Ambisonics (HOA) is an attractive format for immersive spatial audio, but its high channel count and directional properties present significant audio coding challenges.
The widely deployed Opus audio codec supports HOA through Channel Mapping Families~2 and~3 (RFC~8486), which code and transmit every Ambisonics channel, so the bitrate grows quadratically with the spherical harmonic order.
We introduce SEMAC, which instead represents an Ambisonics signal by a quantized full-order directional energy envelope together with fewer CELT-coded transport channels.
CELT separates a per-band energy envelope from the signal shape and restores it after quantization, and SEMAC generalizes this principle to directional signals.
Since the reduced transport spans only a lower-dimensional subspace, the decoder re-imposes the envelope observed at the encoder input, restoring the directional energy distribution at full order.
A MUSHRA test on fifth-order material compares SEMAC against Opus CMF~3 in matched configurations at 512 and 768~kbit/s.
SEMAC with 16 of 36 transport channels is rated above full-order CMF~3 at both rates, and at equal transport channel count above order truncation.
The proposed SEMAC strategy of coding fewer transport channels with the directional envelope is found perceptually preferable to coding all HOA channels at low precision.
\end{abstract}


\section{Introduction}
\label{sec:introduction}

Ambisonics represents a sound field by spherical-harmonic (SH) coefficients and supports reproduction that is independent of both the capture modality and a particular loudspeaker layout, including binaural playback~\cite{zotter2019Ambisonics}.
Its directional resolution increases with the Ambisonics order~$N$, but so does the channel count, $C=(N{+}1)^2$.
Conventional channel-wise coding therefore requires an aggregate rate that grows quadratically with the order when the rate per audio waveform channel is held fixed, which quickly becomes infeasible in practice.
The widely established Opus audio codec supports HOA through Channel Mapping Families (CMFs)~2 and~3~\cite{rfc6716,rfc8486}.
CMF~2 codes the Ambisonics channels independently.
CMF~3 instead applies a fixed linear transform, i.e.\ beamforming, whose outputs are arranged for stereo-coupled Opus coding and are mapped back to HOA channels at the decoder, which has been shown to be perceptually beneficial~\cite{rudzki2019}.
Both code all $C$ audio signals.

Existing approaches aim to reduce the HOA channel count by using inter-channel transforms or parametric descriptions of the sound field.
Signal-adaptive transforms remove inter-channel redundancy explicitly, using general multichannel decorrelation such as principal component or covariance-based transforms (see e.g.\ \cite{YangKLT2003}), which have also been adapted to HOA signals~\cite{petersSMPTE2015,MPEG-H_2017,zamani2017svd,namazi2024svd}.
Parametric spatial audio coding, on the other hand, describes the sound field by high-level acoustic parameters~\cite{pulkki2007dirac,politis2015hodirac,politis2018compass}, such as the number of sources, their directions, or the diffuseness of the sound field.
Parametric HOA coders~\cite{hold2024hoac,Weckbecker2025IVAS,hold2025ivashoa,MPEG-H_2017} then transport few audio waveform channels together with parameter metadata, but their reconstruction quality generally depends on the estimated scene parameters.

While the human auditory system's spatial processing is more complex than mere energy analysis per direction,
including analysis of interaural time-, level-, and correlation cues \cite{blauert1996}, the directional energy composition of a sound field is a key perceptual feature, and SEMAC is built on the premise that preserving it should be prioritized during coding.
SEMAC therefore codes a full-order directional energy envelope separately from the audio waveform transport, where the latter may then be reduced in dimensionality.
This follows the organizing principle of CELT~\cite{valin2013celt}, which separately codes a band envelope and a normalized within-band shape and restores the envelope after quantization.
SEMAC applies the analogous operation across directions for every frequency band, thereby generalizing the core design principle of CELT to the directional domain.

This work makes three contributions.
First, it reduces the number of \emph{transport channels} while transmitting the full-order \emph{directional energy envelope} as side information.
Second, it recovers full-order HOA by applying that envelope to a reduced-order reconstruction, thereby matching the sampled directional magnitudes observed at the input before coding.
Third, it evaluates fifth-order SEMAC against fifth-order Opus CMF~3 in closely matched configurations, isolating the perceptual benefit of restoring the full-order directional envelope from a reduced-order HOA signal.

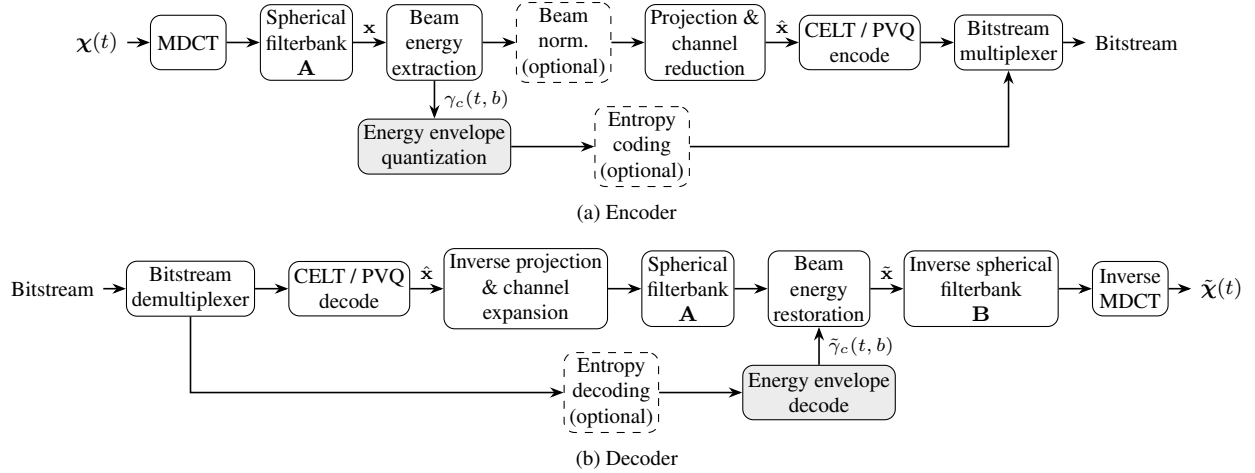
\begin{figure*}[ht]
  \centering
  \begin{tikzpicture}[
    block/.style={draw, rounded corners, minimum width=1.0cm, minimum height=0.72cm,
                  align=center, font=\footnotesize, inner sep=2pt},
    optblock/.style={block, dashed},
    sideblock/.style={block, fill=gray!15},
    arrow/.style={-{Stealth[length=1.8mm]}, semithick},
    lab/.style={font=\scriptsize, midway},
    node distance=0.4cm and 0.44cm]

    \node[font=\footnotesize] (ein) {$\boldsymbol{\chi}(t)$};
    \node[block, right=0.3cm of ein] (mdct) {MDCT};
    \node[block, right=of mdct] (spfb) {Spherical\\filterbank\\$\mathbf{A}$};
    \node[block, right=of spfb] (energy) {Beam\\energy\\extraction};
    \node[optblock, right=of energy] (norm) {Beam\\norm.\\(optional)};
    \node[block, right=of norm] (reduce) {Projection \&\\channel\\reduction};
    \node[block, right=of reduce] (celt) {CELT / PVQ\\encode};
    \node[block, right=of celt] (mux) {Bitstream\\multiplexer};
    \node[font=\footnotesize, right=0.3cm of mux] (eout) {Bitstream};
    \node[sideblock, below=0.5cm of energy] (envq) {Energy envelope\\quantization};
    \node[optblock, right=1.1cm of envq] (entc) {Entropy\\coding\\(optional)};

    \draw[arrow] (ein) -- (mdct);
    \draw[arrow] (mdct) -- (spfb);
    \draw[arrow] (spfb) -- (energy) node[lab, above] {$\mathbf{x}$};
    \draw[arrow] (energy) -- (norm);
    \draw[arrow] (norm) -- (reduce);
    \draw[arrow] (reduce) -- (celt) node[lab, above] {$\hat{\mathbf{x}}$};
    \draw[arrow] (celt) -- (mux);
    \draw[arrow] (mux) -- (eout);
    \draw[arrow] (energy) -- (envq) node[lab, right] {$\gamma_c(t,b)$};
    \draw[arrow] (envq) -- (entc);
    \draw[arrow] (entc) -| (mux);
  \end{tikzpicture}

  {\footnotesize (a) Encoder}

  \vspace{3mm}
  \begin{tikzpicture}[
    block/.style={draw, rounded corners, minimum width=1.0cm, minimum height=0.72cm,
                  align=center, font=\footnotesize, inner sep=2pt},
    optblock/.style={block, dashed},
    sideblock/.style={block, fill=gray!15},
    arrow/.style={-{Stealth[length=1.8mm]}, semithick},
    lab/.style={font=\scriptsize, midway},
    node distance=0.4cm and 0.44cm]

    \node[font=\footnotesize] (din) {Bitstream};
    \node[block, right=0.3cm of din] (demux) {Bitstream\\demultiplexer};
    \node[block, right=of demux] (celtd) {CELT / PVQ\\decode};
    \node[block, right=of celtd] (expand) {Inverse projection\\\& channel\\expansion};
    \node[block, right=of expand] (spfbd) {Spherical\\filterbank\\$\mathbf{A}$};
    \node[block, right=of spfbd] (restore) {Beam\\energy\\restoration};
    \node[block, right=of restore] (ispfb) {Inverse spherical\\filterbank\\$\mathbf{B}$};
    \node[block, right=of ispfb] (imdct) {Inverse\\MDCT};
    \node[font=\footnotesize, right=0.3cm of imdct] (dout) {$\tilde{\boldsymbol{\chi}}(t)$};
    \node[sideblock, below=0.5cm of restore] (envdec) {Energy envelope\\decode};
    \node[optblock, left=1.1cm of envdec] (entd) {Entropy\\decoding\\(optional)};

    \draw[arrow] (din) -- (demux);
    \draw[arrow] (demux) -- (celtd);
    \draw[arrow] (celtd) -- (expand) node[lab, above] {$\hat{\mathbf{x}}$};
    \draw[arrow] (expand) -- (spfbd);
    \draw[arrow] (spfbd) -- (restore);
    \draw[arrow] (restore) -- (ispfb) node[lab, above] {$\tilde{\mathbf{x}}$};
    \draw[arrow] (ispfb) -- (imdct);
    \draw[arrow] (imdct) -- (dout);
    \draw[arrow] (demux.south) |- (entd.west);
    \draw[arrow] (entd) -- (envdec);
    \draw[arrow] (envdec) -- (restore) node[lab, right] {$\tilde{\gamma}_c(t,b)$};
  \end{tikzpicture}

  {\footnotesize (b) Decoder}
  \caption{SEMAC signal flow. (a)~The spherical filterbank transforms SH coefficients into directional beams, from which the per-band directional energy envelope is extracted as side information; the signal is then reduced to fewer transport channels (beam selection in the reported configuration) and coded with CELT. (b)~The decoded transport is expanded to the original order and beamformed by~$\mathbf{A}$; each beam is band-normalized and rescaled by the decoded envelope before the inverse filterbank~$\mathbf{B}$ returns to the SH domain.}
  \label{fig:pipeline}
\end{figure*}

\section{SEMAC Codec}
\label{sec:method}

\subsection{Directional Analysis and Envelope}

Fig.~\ref{fig:pipeline} shows the encoder and decoder signal flow.
Let $\boldsymbol{\chi}(t,k)\in\mathbb{R}^{C}$ contain the MDCT-domain SH coefficients in frame~$t$ and frequency bin~$k$.
The spherical harmonic domain (SHD) signal is directionally decomposed to $D$ beams,
\begin{equation}
  \mathbf{x}(t,k)=\mathbf{A}\boldsymbol{\chi}(t,k),\qquad
  \tilde{\boldsymbol{\chi}}(t,k)=\mathbf{B}\tilde{\mathbf{x}}(t,k),
  \label{eq:filterbank}
\end{equation}
where $\mathbf{A}\in\mathbb{R}^{D\times C}$ is the analysis matrix and $\mathbf{B}=\mathbf{A}^{\dagger}$ its synthesis pseudoinverse.
Throughout, $\hat{\mathbf{x}}$ marks a transport quantity and $\tilde{\mathbf{x}}$ a corresponding restored output of $\mathbf{x}$.
We construct $\mathbf{A}$ from max-$r_E$ beamformers evaluated on a critically sampled spherical grid~\cite{hold2021spatial},
\begin{equation}
  \mathbf{A}=\mathbf{Y}\,\mathrm{diag}(\mathbf{c}_N),\qquad
  \mathbf{B}=(\mathbf{A}^{\mathsf T}\mathbf{A})^{-1}\mathbf{A}^{\mathsf T},
  \label{eq:spfb}
\end{equation}
where the rows of $\mathbf{Y}\in\mathbb{R}^{D\times C}$ are real SH basis functions at the beam directions and $\mathbf{c}_N$ contains the modal weights repeated over degrees of each order.
The directional analysis--synthesis matrix pair can be interpreted as a spherical filterbank.
With full column rank, $\mathbf{B}\mathbf{A}=\mathbf{I}_C$ before coding.
The derivation in this paper uses orthonormal N3D SH convention; note that input and output are converted to and from the SN3D AmbiX convention used by the implementation.
The reported fifth-order experiments use $C=D=36$, with quasi-uniform spherical sampling on a critical sampling grid from \cite{sloan2004extremal}.

For every beam channel~$c$ and CELT frequency band~$b$ (which approximate critical bands), the encoder extracts the band magnitude
\begin{equation}
  \gamma_c(t,b)=
  \left(\sum_{k\in\mathcal{B}_b}|x_c(t,k)|^2+\epsilon\right)^{1/2},
  \label{eq:envelope}
\end{equation}
where $\mathcal{B}_b$ is the set of MDCT bins in band~$b$ and $\epsilon>0$ is a numerical floor.
The collection $\boldsymbol{\gamma}(t)$ is the directional energy envelope, sampled at the $D$ beam directions and stored as per-band magnitudes.
Unlike a source model, it records no explicit direction-of-arrival or source-count estimate.
The aggregate rate is
$R_{\mathrm{tot}}=R_{\mathrm{CELT}}(C_{\mathrm{tp}})+R_{\gamma}$, 
so reducing the number of transport channels $C_{\mathrm{tp}}$ trades off against the side-information rate $R_{\gamma}$.

\subsection{Envelope Coding}
\label{sec:envelope_coding}

A direct representation contains $D\times B$ magnitudes per frame, where, in the presented study, $D=36$ directions, $B=21$ bands, and $50$ frames/s.
SEMAC therefore reuses CELT's energy-coding structure rather than transmitting independent scalar values.
For each beam, a coarse log-energy is predicted from the preceding frame and the preceding band, and the prediction residual is entropy coded on a $6\,\mathrm{dB}$ grid.
Fine bits then refine selected bands.
This construction exploits temporal and spectral correlation while retaining independent directional targets; the decoder reconstructs the same quantized envelope $\tilde{\boldsymbol{\gamma}}$ before decoding the normalized shapes.
The number of fine bits follows the available per-transport-channel rate: low-rate modes reduce envelope precision, whereas higher-rate modes retain the full refinement, which prevents the side information from dominating when the transport is bit-starved.

The side stream scales with the input signal order rather than the transport order; consequently, increasing the input order raises $R_\gamma$ even if $C_{\mathrm{tp}}$ is fixed.
The useful operating point is therefore determined jointly by SH order, transport dimension, envelope precision, and aggregate rate, which the experiments expose by comparing $C_{\mathrm{tp}}\in\{4,9,16\}$ under a fixed fifth-order envelope.

\subsection{Reduced Audio Channel Transport}

Our objective is to find a coding domain that (i) allows the number of transport channels to be scaled while preserving the relevant signal components, and (ii) offers favorable coding properties such as decorrelation and energy compaction.
This work uses signal-independent transforms, so that the transport channels are either SHD signals or selected beamformer outputs; signal-dependent alternatives such as the SVD are drop-in replacements within the same framework.

Let $\mathbf{T}\in\mathbb{R}^{C_{\mathrm{tp}}\times C}$ be a projection onto $C_{\mathrm{tp}}<C$ transport channels and $\mathbf{U}\in\mathbb{R}^{C\times C_{\mathrm{tp}}}$ the corresponding reconstruction.
The transport and its reconstruction are
\begin{equation}
  \hat{\mathbf{x}}(t,k)=\mathbf{T}\boldsymbol{\chi}(t,k),\qquad
  \hat{\boldsymbol{\chi}}(t,k)=\mathbf{U}\hat{\mathbf{x}}(t,k),
  \label{eq:transport}
\end{equation}
extracting the transport channels $\hat{\mathbf{x}}$.
The rank of $\mathbf{U}$ is at most $C_{\mathrm{tp}}$, so $\hat{\boldsymbol{\chi}}$ is confined to a $C_{\mathrm{tp}}$-dimensional subspace; in practice the reconstruction before envelope restoration therefore corresponds to a signal of reduced spherical order.
Both $\mathbf{T}$ and $\mathbf{U}$ are fixed in the reported configuration and require no matrix side information, whereas a signal-adaptive choice would have to transmit its basis.
For $C_{\mathrm{tp}}=C$ and $D=C$, setting $\mathbf{T}=\mathbf{A}$ and $\mathbf{U}=\mathbf{B}$ recovers the mechanism of Opus CMF~3.

SEMAC biases the CELT bitrate allocation to reflect the transport channel band energy, using zero-sum offsets that preserve the total per-band bitrate.
Because encoder and decoder derive them from identically decoded CELT band magnitudes, no allocation side information is required.

A key proposal of SEMAC is that $\mathbf{T}$ may extract either SHD signals or directionally selective signals, depending on the content, and may be time- and frequency-dependent.
While the SHD domain inherently decorrelates diffuse scenes and, for typical scenes, concentrates energy in the lower orders, coding SHD signals directly may spread the CELT coding noise across directions.
Directionally selective transport channels, on the other hand, represent scenes with few sources (e.g., plane waves) very compactly, and coding them keeps the coding noise directionally coherent with the signal.

The evaluation in this paper uses beam-domain transport throughout, which matches the transport domain of Opus CMF~3 and thereby isolates the contribution of SEMAC's directional energy matching more clearly.
The encoder may optionally whiten the beams by $\gamma_c(t,b)$ before reduction, which is not applied here.
CELT still normalizes each transport channel per band as part of its own shape coding.

\subsection{Envelope-Guided Reconstruction}

Because of the lossy coding and transport paradigm, the directional energy envelope of the decoded signal generally differs from that of the input.
Therefore, after expansion, the decoder analyzes the decoded $\hat{\boldsymbol{\chi}}$ with $\mathbf{A}$, i.e.\ with the same full directional resolution as during encoding.
For $k\in\mathcal{B}_b$, it corrects the envelope measured at the decoder, $\hat{\gamma}_c(t,b)$, obtained by applying Eq.~\eqref{eq:envelope} to $\hat{\mathbf{x}}$, towards the decoded input envelope $\tilde{\gamma}_c(t,b)$:
\begin{equation}
  \tilde{x}_c(t,k)
    =\frac{\tilde{\gamma}_c(t,b)}{\hat{\gamma}_c(t,b)}{\hat{x}_c(t,k)},
  \label{eq:restore}
\end{equation}
where the output is projected back to the SHD with $\tilde{\boldsymbol{\chi}}=\mathbf{B}\tilde{\mathbf{x}}$.
The rescaling sets each beam's band magnitude to its transmitted target, so the decoder imposes the input directional energy envelope on the rank-limited reconstruction of Eq.~\eqref{eq:transport}.
This restores the full-order directional energy distribution and may be interpreted as an energy-guided upmix from the reduced transport order.

\subsection{Relation to CELT and CMF~3}

CMF~3 applies a square, fixed projection before Opus and its inverse after decoding, which organizes directional signals for stereo coupling but retains all $C$ waveform dimensions.
SEMAC adds two operations around the audio waveform codec: reduction to $C_{\mathrm{tp}}<C$ dimensions, and spherical full-order envelope restoration after expansion, whereas CMF~3 can only restore energy per individual channel.
The two mechanisms are complementary, in that CELT controls the spectral magnitude of each transport channel, whereas SEMAC controls directional band magnitudes after full-order reconstruction.

\section{Perceptual Evaluation}
\label{sec:evaluation}

\subsection{Setup}

The reported experiments use 48-kHz fifth-order Ambisonics signals (36 channels).
Participants rated overall audio quality following the MUSHRA paradigm~\cite{itu2015mushra}, with a hidden reference and a 3.5-kHz low-pass anchor present in every trial.
This work evaluates a test set containing seven items that were found \textit{critical} to spatial audio coding in a pre-selection process, spanning sparse, dense, and reverberant scenes.
Each item was presented at aggregate bitrates of 512 and 768~kbit/s, yielding 14 trials per listener.
All reported codecs are forked from Opus v1.6.1 and use 20-ms frames.
SEMAC is evaluated with 4, 9, or 16 transport channels, configured for beam-domain transport in this comparison.
The baseline is a full-order, 36-channel variant of Opus CMF~3, configured with identical mixing and demixing matrices, i.e.\ beam-domain transport, and with CELT coding only, in order to isolate the contribution of the proposed directional energy matching.
The corresponding reduced-transport-channel CMF~3 conditions (SH1, SH2), obtained by input-order truncation, are included for comparison.
Note that the output for the latter is therefore not fifth-order, whereas all other conditions are.
All outputs are rendered binaurally using magnitude least-squares (MagLS) decoding~\cite{schoerkhuber2018magls} with KU100 HRTFs.
They are scored against the uncoded fifth-order signal rendered through the same path, which also serves as the hidden reference.

\subsection{Results and Discussion}

\begin{figure}[t]
  \centering
  \includegraphics[width=\columnwidth]{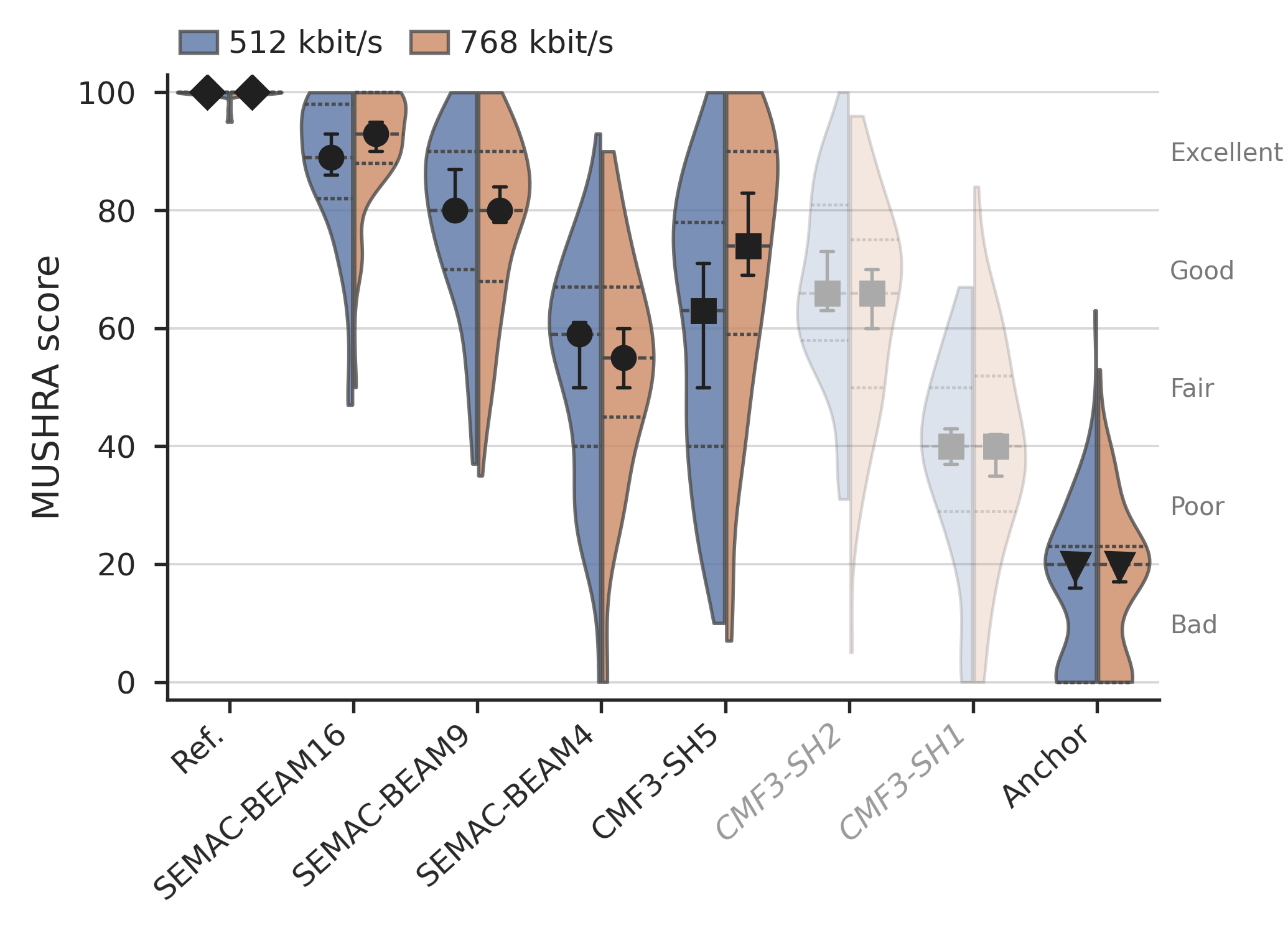}
  \caption{Violin plots with quartiles, and medians with bootstrapped 95\,\% confidence intervals of MUSHRA ratings for fifth-order input at 512 and 768~kbit/s.
  SEMAC is compared against Opus CMF~3 with matched parameters.
  Labels give the SEMAC transport-channel count, where the decoder recovers full order.
  For CMF~3, only CMF3-SH5 transports the full order, whereas CMF3-SH2 and CMF3-SH1 apply order truncation and are shown greyed.
  }
  \label{fig:mushra}
\end{figure}

Fifteen listeners completed the test, out of which four were removed by post-screening (rating the hidden reference below~90 in at least one trial), leaving 11 subjects.
Figure~\ref{fig:mushra} shows the resulting ratings.
The hidden reference was rated at ceiling and the anchor received consistently low scores, indicating that listeners used the scale as intended.

For SEMAC, ratings increase monotonically with the number of transport channels, and the 16-channel configuration reaches the \textit{Excellent} range.
For the 16-channel configuration, 768~kbit/s suggests improved performance over 512~kbit/s, whereas the 9 and 4-channel configurations perform similarly between the two bitrates.
This suggests that for greatly reduced transport channels, 512~kbit/s may already suffice, and the upper quality limit is bound by the number of transport channels, rather than the available bitrate.

The baseline behaves differently.
Full-order CMF~3 (CMF3-SH5, 36 transport channels) gains no clear advantage over CMF3-SH2 with only 9 transport channels, despite coding four times as many channels, with a large spread of the ratings at both bitrates.
CMF3-SH5 also improves clearly with bitrate.
Both observations are consistent with an insufficient per-channel bitrate: allocating across all 36 channels starves each of them, so the full-order baseline cannot convert its higher channel count into higher quality.
The naive order truncation (CMF3-SH2 and CMF3-SH1) shows clear quality deterioration, leaving both well below the reference.

Two comparisons isolate the contribution of the proposed envelope restoration.
First, at equal transport channel count SEMAC is rated above the truncated baseline, compare e.g. CMF3-SH2 and SEMAC-BEAM9 ($C_{\mathrm{tp}} = 9$).
Second, SEMAC with 16 transport channels exceeds full-order CMF~3 with 36 channels at both bitrates, with higher medians and a tighter spread of the quality ratings.

Two interpretations follow.
First, at the tested rates, allocating part of the bitstream to a directional energy envelope is more effective than coding every full-order waveform audio channel at low precision.
Second, restoring the envelope at the decoder is perceptually effective, since SEMAC outperforms naive transport channel reduction by order truncation at equal transport channel count.

We also identified limitations of the current implementation.
A denser spherical grid samples the envelope more finely but raises its rate: at fifth order the envelope quantization can consume more than $100\,\mathrm{kbit/s}$, which may starve the core coder prematurely, so the optimal split between waveform and metadata rate needs further investigation.
Fewer transport channels save waveform rate but discard more within-band directional structure, relying more heavily on the metadata.
The decoder can reapply the magnitudes sampled at discrete directions, but cannot reconstruct phase relationships or directional detail absent from the reduced transport.

Although the codec was continually assessed with informal listening and objective quality models over a larger test set, including loudspeaker playback, a more comprehensive evaluation and comparison is left for future work.

\section{Conclusion}
\label{sec:conclusion}

SEMAC augments mono and stereo CELT audio waveform coding with a quantized directional energy envelope measured at the encoder input.
Because the reduced transport spans only a lower-dimensional subspace, and thereby introduces directional error, the decoder re-imposes this envelope to restore the directional energy distribution.
In a MUSHRA test on fifth-order material, SEMAC with 16 of 36 transport channels was rated above full-order Opus CMF~3 at both tested bitrates, and it exceeded naive order truncation at equal transport channel count.
This indicates that spending part of the bit allocation on the directional energy envelope is preferable to coding all HOA channels at low precision.

\section{Acknowledgements}
The authors thank Katarzyna Sochaczewska and Tomasz Rudzki for creating and providing the listening test items, and all test participants.

\vfill\pagebreak

\bibliographystyle{IEEEbib}
\bibliography{references}

@techreport{itu2015mushra,
  author      = {{International Telecommunication Union}},
  title       = {Method for the Subjective Assessment of Intermediate Quality Level of Audio Systems},
  institution = {ITU-R},
  type        = {Recommendation},
  number      = {BS.1534-3},
  year        = {2015}
}

@book{blauert1996,
  author    = {Blauert, Jens},
  title     = {Spatial Hearing: The Psychophysics of Human Sound Localization},
  publisher = {MIT Press},
  year      = {1996},
  doi       = {10.7551/mitpress/6391.001.0001}
}

@inproceedings{hold2021spatial,
  author    = {Hold, Christoph and Schlecht, Sebastian J and Politis, Archontis and Pulkki, Ville},
  title     = {Spatial filter bank in the spherical harmonic domain: Reconstruction and application},
  booktitle = {Proceedings of the IEEE Workshop on Applications of Signal Processing to Audio and Acoustics (WASPAA)},
  pages     = {361--365},
  year      = {2021},
  doi       = {10.1109/waspaa52581.2021.9632709}
}

@inproceedings{namazi2024svd,
  author    = {Namazi, M. and Hirvonen, T. and Bharitkar, S. and Souza-Blanes, E.},
  title     = {Pre- and Post-Processing of Ambisonics Signals for Legacy Audio Codecs},
  booktitle = {Proceedings of the 156th AES Convention},
  year      = {2024}
}

@techreport{rfc6716,
  author      = {Valin, Jean-Marc and Vos, Koen and Terriberry, Timothy B.},
  title       = {Definition of the Opus Audio Codec},
  institution = {Internet Engineering Task Force},
  type        = {RFC},
  number      = {6716},
  year        = {2012},
  doi         = {10.17487/RFC6716},
  url         = {https://www.rfc-editor.org/rfc/rfc6716}
}

@techreport{rfc8486,
  author      = {Jan Skoglund and Michael Graczyk},
  title       = {Ambisonics in an Ogg Opus Container},
  institution = {Internet Engineering Task Force},
  type        = {RFC},
  number      = {8486},
  year        = {2018},
  doi         = {10.17487/RFC8486},
  url         = {https://www.rfc-editor.org/rfc/rfc8486}
}

@inproceedings{rudzki2019,
  author = {Rudzki, Tomasz and Hening, Pierce and Gomez-Lanzaco, Ignacio and Stubbs, Jessica and McKenzie, Thomas and Skoglund, Jan and Murphy, Damian and Kearney, Gavin},
  year = {2019},
  booktitle = {2019 AES International Conference on Immersive and Interactive Audio},
  pages = {},
  title = {Perceptual Evaluation of Bitrate Compressed Ambisonic Scenes in Loudspeaker Based Reproduction}
}

@inproceedings{schoerkhuber2018magls,
  author    = {Schörkhuber, Christian and Zaunschirm, Matthias and Höldrich, Robert},
  title     = {Binaural Rendering of Ambisonic Signals via Magnitude Least Squares},
  booktitle = {Proceedings of DAGA},
  address   = {Munich, Germany},
  year      = {2018}
}

@article{sloan2004extremal,
  author    = {Sloan, Ian H and Womersley, Robert S},
  title     = {Extremal systems of points and numerical integration on the sphere},
  journal   = {Advances in Computational Mathematics},
  volume    = {21},
  number    = {1},
  pages     = {107--125},
  year      = {2004},
  publisher = {Springer},
  doi       = {10.1023/b:acom.0000016428.25905.da}
}

@inproceedings{valin2013celt,
  author    = {Valin, Jean-Marc and Maxwell, Gregory and Terriberry, Timothy B. and Vos, Koen},
  title     = {High-Quality, Low-Delay Music Coding in the Opus Codec},
  booktitle = {Proceedings of the 135th AES Convention},
  year      = {2013}
}

@inproceedings{zamani2017svd,
  author    = {Zamani, S. and Nanjundaswamy, T. and Rose, K.},
  title     = {Frequency-Domain Singular Value Decomposition for Efficient Spatial Audio Coding},
  booktitle = {Proceedings of the IEEE Workshop on Applications of Signal Processing to Audio and Acoustics (WASPAA)},
  pages     = {161--165},
  year      = {2017},
  doi       = {10.1109/waspaa.2017.8170008}
}

@book{zotter2019Ambisonics,
  author    = {Zotter, Franz and Frank, Matthias},
  title     = {Ambisonics: A Practical 3D Audio Theory for Recording, Studio Production, Sound Reinforcement, and Virtual Reality},
  publisher = {Springer},
  year      = {2019},
  doi       = {10.1007/978-3-030-17207-7}
}

@article{pulkki2007dirac,
  author={Pulkki, Ville},
  title={Spatial sound reproduction with directional audio coding},
  journal={Journal of the Audio Engineering Society},
  year={2007},
  affiliation={Laboratory of Acoustics and Audio Signal Processing, Helsinki University of Technology, 02015, Finland},
  descriptors={Acoustique audio; Angle arrivée; Codage; Localisation; Microphone; Méthode domaine temps fréquence; Reproduction son; Représentation spatiale; Réponse impulsion; Réponse transitoire; Synthèse signal},
  subject={Transducteurs et dispositif pour la génération et la reproduction du son[001B40C38]},
  issn={0004-7554},
  document_type={Article},
  inist_number={18877676}
}

@ARTICLE{politis2015hodirac,
  author={Politis, Archontis and Vilkamo, Juha and Pulkki, Ville},
  journal={IEEE Journal of Selected Topics in Signal Processing},
  title={Sector-Based Parametric Sound Field Reproduction in the Spherical Harmonic Domain},
  year={2015},
  volume={9},
  number={5},
  pages={852-866},
  doi={10.1109/JSTSP.2015.2415762}
}

@inproceedings{politis2018compass,
  author={Politis, Archontis and Tervo, Sakari and Pulkki, Ville},
  booktitle={2018 IEEE International Conference on Acoustics, Speech and Signal Processing (ICASSP)},
  title={COMPASS: Coding and Multidirectional Parameterization of Ambisonic Sound Scenes},
  year={2018},
  volume={},
  number={},
  pages={6802-6806},
  doi={10.1109/ICASSP.2018.8462608}
}

@inproceedings{Weckbecker2025IVAS,
  author={Weckbecker, Dominik and Brown, Stefanie and Torres, Juan and Multrus, Markus and Tamarapu, Archit and Fuchs, Guillaume},
  booktitle={2025 IEEE International Conference on Acoustics, Speech and Signal Processing (ICASSP)},
  title={Ambisonics Coding in IVAS: A Hybrid SPAR and DirAC System},
  year={2025},
  volume={},
  number={},
  pages={1-5},
  doi={10.1109/ICASSP49660.2025.10888651}
}

@inproceedings{hold2025ivashoa,
  author={Hold, Christoph and Weckbecker, Dominik and Fuchs, Guillaume and Multrus, Markus and Tyagi, Rishabh and Brown, Stefanie and Torres, Juan and Bruhn, Stefan},
  booktitle={2025 33rd European Signal Processing Conference (EUSIPCO)},
  title={Coding Higher Order Ambisonics in {3GPP IVAS} – Scaling Parametric Audio Coding to Higher Bitrates},
  year={2025},
  volume={},
  number={},
  pages={196-200},
  doi={10.23919/EUSIPCO63237.2025.11226809}
}

@inproceedings{hold2024hoac,
  author={Hold, Christoph and McCormack, Leo and Politis, Archontis and Pulkki, Ville},
  booktitle={ICASSP 2024 - 2024 IEEE International Conference on Acoustics, Speech and Signal Processing (ICASSP)},
  title={Perceptually-Motivated Spatial Audio Codec for Higher-Order Ambisonics Compression},
  year={2024},
  volume={},
  number={},
  pages={1121-1125},
  doi={10.1109/ICASSP48485.2024.10447577}
}

@ARTICLE{YangKLT2003,
  author={Dai Yang and Hongmei Ai and Kyriakakis, C. and Kuo, C.-C.J.},
  journal={IEEE Transactions on Speech and Audio Processing},
  title={High-fidelity multichannel audio coding with Karhunen-Loeve transform},
  year={2003},
  volume={11},
  number={4},
  pages={365-380},
  doi={10.1109/TSA.2003.814375}}

@inproceedings{petersSMPTE2015,
  author    = {Nils Peters and Deep Sen and Moo-Young Kim and Oliver Wuebbolt and S. Merrill Weiss},
  title     = {Scene-based Audio Implemented with Higher Order Ambisonics (HOA)},
  booktitle = {SMPTE 2015 Annual Technical Conference and Exhibition},
  year      = {2015},
  publisher = {Society of Motion Picture and Television Engineers},
  doi       = {10.5594/M001651}
}

@ARTICLE{MPEG-H_2017,
  author={Bleidt, Robert L. and others},
  journal={IEEE Transactions on Broadcasting},
  title={Development of the {MPEG-H TV} Audio System for {ATSC} 3.0},
  year={2017},
  volume={63},
  number={1},
  pages={202-236},
  doi={10.1109/TBC.2017.2661258}}

\end{document}